\documentclass[twocolumn]{aastex631}

\newcommand{\lenstronomy}{\texttt{lenstronomy}}
\newcommand{\galight}{\texttt{galight}}

\usepackage{indentfirst}
\usepackage{appendix}
\usepackage{makecell}
\usepackage{tabu}
\usepackage{float}
\usepackage{xcolor}
\usepackage{enumitem}
\usepackage{amsmath}

\def\ss{\textit{$\rm S\acute{e}rsic$}}

\shortauthors{Silverman et al.}

\begin{document}


\title{Resolving galactic-scale obscuration of X-ray AGN at $z\gtrsim 1$ with COSMOS-Web}


\author{John D. Silverman}
\affiliation{Kavli Institute for the Physics and Mathematics of the Universe (Kavli IPMU, WPI), UTIAS, Tokyo Institutes for Advanced Study, University of Tokyo, Chiba, 277-8583, Japan}
\affiliation{Department of Astronomy, School of Science, The University of Tokyo, 7-3-1 Hongo, Bunkyo, Tokyo 113-0033, Japan}
\affiliation{Center for Data-Driven Discovery, Kavli IPMU (WPI), UTIAS, The University of Tokyo, Kashiwa, Chiba 277-8583, Japan}
\author[0000-0002-1047-9583]{Vincenzo Mainieri}
\affiliation{European Southern Observatory, Karl-Schwarzschild-Strasse 2, D-85748, Garching bei Munchen, Germany}
\author{Xuheng Ding}
\affiliation{Kavli Institute for the Physics and Mathematics of the Universe (Kavli IPMU, WPI), UTIAS, Tokyo Institutes for Advanced Study, University of Tokyo, Chiba, 277-8583, Japan}
\author{Daizhong Liu}
\affiliation{Max-Planck-Institut f\"ur extraterrestrische Physik, Giessenbachstraße 1, D-85748 Garching, Germany}
\author{Knud Jahnke}
\affiliation{Max Planck Institute for Astronomy, Königstuhl 17, D-69117 Heidelberg, Germany}
\author{Michaela Hirschmann}
\affiliation{Institute for Physics, Laboratory for Galaxy Evolution and Spectral modelling, Ecole Polytechnique Federale de Lausanne, Observatoire de Sauverny, Chemin Pegasi 51, 1290 Versoix, Switzerland}
\author{Jeyhan Kartaltepe}
\affiliation{Laboratory for Multiwavelength Astrophysics, School of Physics and Astronomy, Rochester Institute of Technology, 84 Lomb Memorial Drive, Rochester, NY 14623, USA}
\author{Erini Lambrides}
\affiliation{NASA Goddard Space Flight Center, Code 662, Greenbelt, MD, 20771, USA}
\author{Masafusa Onoue}
\affiliation{Kavli Institute for Astronomy and Astrophysics, Peking University, Beijing 100871, China}
\affiliation{Kavli Institute for the Physics and Mathematics of the Universe (Kavli IPMU, WPI), UTIAS, Tokyo Institutes for Advanced Study, University of Tokyo, Chiba, 277-8583, Japan}
\author{Benny Trakhtenbrot}
\affiliation{School of Physics and Astronomy, Tel Aviv University, Tel Aviv 69978, Israel}
\author{Eleni Vardoulaki}
\affiliation{Thüringer Landessternwarte, Sternwarte 5, D-07778 Tautenburg, Germany}
\author{Angela Bongiorno}
\affiliation{INAF, Osservatorio Astronomico di Roma, Via Frascati 33, 00078 Monte Porzio Catone, Italy}
\author{Caitlin Casey}
\affiliation{The University of Texas at Austin, 2515 Speedway Blvd Stop C1400, Austin, TX 78712, USA}
\author{Francesca Civano}
\affiliation{Center for Astrophysics — Harvard \& Smithsonian, Cambridge, MA 02138, USA}
\author{Andreas Faisst}
\affiliation{Caltech/IPAC, 1200 E.\ California Blvd., Pasadena, CA 91125, USA}
\author{Maximilien Franco}
\affiliation{The University of Texas at Austin, 2515 Speedway Blvd Stop C1400, Austin, TX 78712, USA}
\author{Steven Gillman}
\affiliation{Cosmic Dawn Center (DAWN), Denmark}
\affiliation{DTU-Space, Technical University of Denmark, Elektrovej 327, 2800 Kgs.\ Lyngby, Denmark}
\author{Ghassem Gozaliasl}
\affiliation{Department of Physics, University of Helsinki, P.O.\ Box 64, FI-00014 Helsinki, Finland}
\author{Christopher C. Hayward}
\affiliation{Center for Computational Astrophysics, Flatiron Institute, 162 Fifth Avenue, New York, NY 10010, USA}
\author[0000-0002-6610-2048]{Anton M. Koekemoer}
\affiliation{Space Telescope Science Institute, 3700 San Martin Dr., Baltimore, MD 21218, USA}
\author{Vasily Kokorev}
\affiliation{Kapteyn Astronomical Institute, University of Groningen, P.O. Box 800, 9700AV Groningen, The Netherlands}
\author{Georgios Magdis}
\affiliation{Cosmic Dawn Center (DAWN), Jagtvej 128, 2200 Copenhagen N, Denmark}
\affiliation{DTU-Space, Technical University of Denmark, Elektrovej 327, 2800 Kgs.\ Lyngby, Denmark}
\affiliation{Niels Bohr Institute, University of Copenhagen, Jagtvej 128, 2200 Copenhagen N, Denmark}
\author{Stefano Marchesi}
\affiliation{Dipartimento di Fisica e Astronomia, Università degli Studi di Bologna, via Gobetti 93/2, 40129 Bologna, Italy}
\affiliation{Department of Physics and Astronomy, Clemson University,  Kinard Lab of Physics, Clemson, SC 29634, USA}
\affiliation{INAF - Osservatorio di Astrofisica e Scienza dello Spazio di Bologna, Via Piero Gobetti, 93/3, 40129, Bologna, Italy}
\author{Robert Michael Rich}
\affiliation{Department of Physics and Astronomy, UCLA, PAB 430 Portola Plaza, Box 951547, Los Angeles, CA 90095, USA}
\author{Martin Sparre}
\affiliation{Institut für Physik und Astronomie, Universität Potsdam, Karl-Liebknecht-Str.~24/25, D-14476 Golm, Germany}
\affiliation{Leibniz-Institut für Astrophysik Potsdam (AIP), An der Sternwarte 16, D-14482 Potsdam, Germany}
\author{Hyewon Suh}
\affiliation{Gemini Observatory/NSF’s NOIRLab, 670 N. A’ohoku Place, Hilo, Hawaii, 96720, USA}
\author{Takumi Tanaka}
\affiliation{Kavli Institute for the Physics and Mathematics of the Universe (Kavli IPMU, WPI), UTIAS, Tokyo Institutes for Advanced Study, University of Tokyo, Chiba, 277-8583, Japan}
\affiliation{Department of Astronomy, School of Science, The University of Tokyo, 7-3-1 Hongo, Bunkyo, Tokyo 113-0033, Japan} 
\author{Francesco Valentino}
\affiliation{European Southern Observatory, Karl-Schwarzschild-Strasse 2, D-85748, Garching bei Munchen, Germany}
\affiliation{Cosmic Dawn Center (DAWN), Denmark}



\begin{abstract}

A large fraction of the accreting supermassive black hole population is shrouded by copious amounts of gas and dust, particularly in the distant ($z\gtrsim1$) Universe. While much of the obscuration is attributed to a parsec-scale torus, there is a known contribution from the larger-scale host galaxy. Using JWST/NIRCam imaging from the COSMOS-Web survey, we probe the galaxy-wide dust distribution in X-ray selected AGN up to $z\sim2$. Here, we focus on a sample of three AGNs with their host galaxies exhibiting prominent dust lanes, potentially due to their edge-on alignment. These represent 27\% (3 out of 11 with early NIRCam data) of the heavily obscured ($N_H>10^{23}$\,cm$^{-2}$) AGN population. With limited signs of a central AGN in the optical and near-infrared, the NIRCam images are used to produce reddening maps $E(B-V)$ of the host galaxies. We compare the mean central value of $E(B-V)$ to the X-ray obscuring column density along the line-of-sight to the AGN ($N_H\sim10^{23-23.5}$\,cm$^{-2}$). We find that the extinction due to the host galaxy is present ($0.6\lesssim E(B-V) \lesssim 0.9$; $1.9 \lesssim A_V 
\lesssim 2.8$) and significantly contributes to the X-ray obscuration at a level of $N_H\sim10^{22.5}$\,cm$^{-2}$ assuming an SMC gas-to-dust ratio which amounts to $\lesssim$\,30\% of the total obscuring column density. These early results, including three additional cases from CEERS, demonstrate the ability to resolve such dust structures with JWST and separate the different circumnuclear and galaxy-scale obscuring structures. 
\end{abstract}









\section{Introduction}

Supermassive black holes at the center of massive galaxies are usually found behind a cloak of gas and dust \citep{Hickox2018}. Understanding the nature of this obscuring medium has long been an area of active research, particularly regarding the inner structure of Active Galactic Nuclei (AGN) and the demographics of the population \citep{Brandt2015,Almeida2017}. On the latter, large samples of obscured AGN across cosmic time \citep[e.g., ][]{Hasinger2008,Merloni2014,Buchner2015,Vito2018,Koss2022} have been constructed from surveys at a multitude of wavelengths such as the X-rays which can penetrate moderate columns of gas and the infrared which represents the reprocessed radiation.

One specific issue is the spatial location of the obscuration. A sub-pc scale torus is the canonical feature that explains several AGN characteristics including their spectral energy distributions (SED), classification, and polarization measurements \citep[e.g.,][]{Antonucci1993, Netzer2015}. Further, the contribution from the host galaxy on larger scales can be present as well \citep[e.g.,][]{Goulding2012}. This is particularly the case if the galaxy is undergoing a gas-rich merger which compresses the gas/dust and increases the central surface density leading to further obscuration \citep{Blecha2018,McKinney2021}. At low redshifts ($z<0.035$), \citet{Malkan1998} report multiple galactic structures including dust lanes based on HST/WFC2 imaging of a large sample of Seyfert galaxies. Many radio-loud AGN at $z<0.5$ with disk-like hosts also exhibit prominant dust lanes \citep{Wu2022}. As a consequence, such galaxy-wide extinction results in the well-known bias of AGN type with host galaxy inclination \citep{Maiolino1995}. Recently, this has been clearly exemplified up to $z\sim0.8$ using a large sample of quasars from SDSS and X-ray AGNs from eROSITA with Subaru/Hyper Suprime-Cam optical imaging of their host galaxies \citep{Li2023}. Similarly, \citet{Baron2016}, from the stacking analysis of $\sim 5000$ SDSS AGNs at z$\sim 0.4$, concluded that they are generally reddened by dust along the line of sight. Such obscuration by the host galaxies of AGN is expected to increase with lookback time due to the rise of the gas fraction of star-forming galaxies \citep{Buchner2017,Gilli2022}.

Now, there are new capabilities to image the host galaxies of AGN with the \textit{James Webb Space Telescope} (JWST) from the near-to-mid infrared to pierce through such obscuring material in high redshift ($z\gtrsim1$) systems. In particular, the Near Infrared Camera (NIRCam) offers the spatial resolution to directly image the galaxy-scale dust distribution up to $z\sim3$. 
Here we utilize the first six pointings from the multi-band COSMOS-Web survey \citep{Casey2022} to demonstrate such capability. We specifically focus on 3 X-ray AGN that exhibit spatially-resolved dust lanes in the bluer NIRCam filters (F115W and F150W). The longer wavelength filters (F277W and F444W) are not as affected by this obscuration, thus providing a penetrating view of the embedded AGN for one case. We perform 2D decomposition (AGN+host) and SED fitting to produce spatially-resolved reddening maps which aid in determining the contribution of galactic extinction, due to dust lanes, to the total X-ray absorption. Throughout this paper we use a Hubble constant of ${\mathrm H}_0 = 70$\,km\,s$^{-1}$\,Mpc$^{-1}$ and cosmological density parameters $\Omega_\mathrm{m} = 0.3$ and $\Omega_\Lambda = 0.7$. 

\section{Data}

\subsection{COSMOS-Web}

COSMOS-Web \citep{Casey2022} is 255-hour treasury program conducted by JWST in its first cycle of observations. The primary science aim is to map the large-scale galaxy distribution beyond $z\sim6$. In addition, there is a wealth of spatially-resolved galaxies at lower redshifts having improved spatial resolution, sensitivity, and longer wavelength coverage as compared to HST and Spitzer. Four filters (F115W, F150W, F277W and F444W) with NIRCam \citep{Rieke2023} are used to reach a 5$\sigma$ point source sensitivity of 27.5--28.2 magnitudes across a contiguous region of 0.54 deg$^2$. This is accomplished for each pointing and filter with a total exposure time of $\sim34$ minutes (see Sec. 2.1 of \cite{Casey2022} for a full description of the observing strategy). In parallel, MIRI imaging \citep{Bouchet2015} in one filter (F770W) is taken over 0.19\,deg$^2$. Here, we utilize the NIRCam imaging from the first 6 pointings (0.02\,deg$^2$) taken on January 5--6, 2023 to study the host galaxy properties of a select sample of obscured X-ray AGN. Details of the image reduction will be presented in Franco et al.\ (in preparation). 

\begin{figure*}
\epsscale{1.1}
\plotone{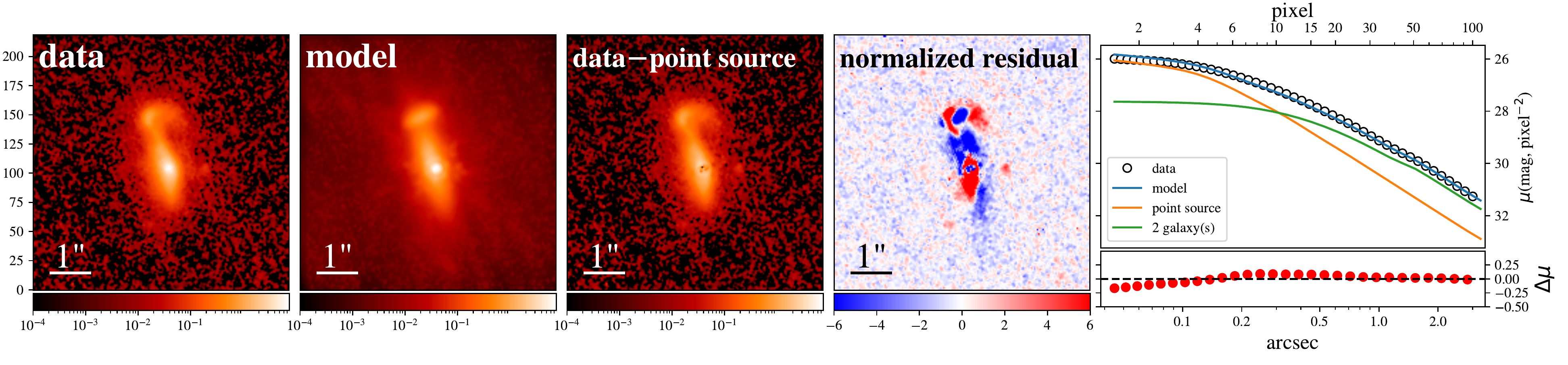}
\epsscale{0.88}
\plotone{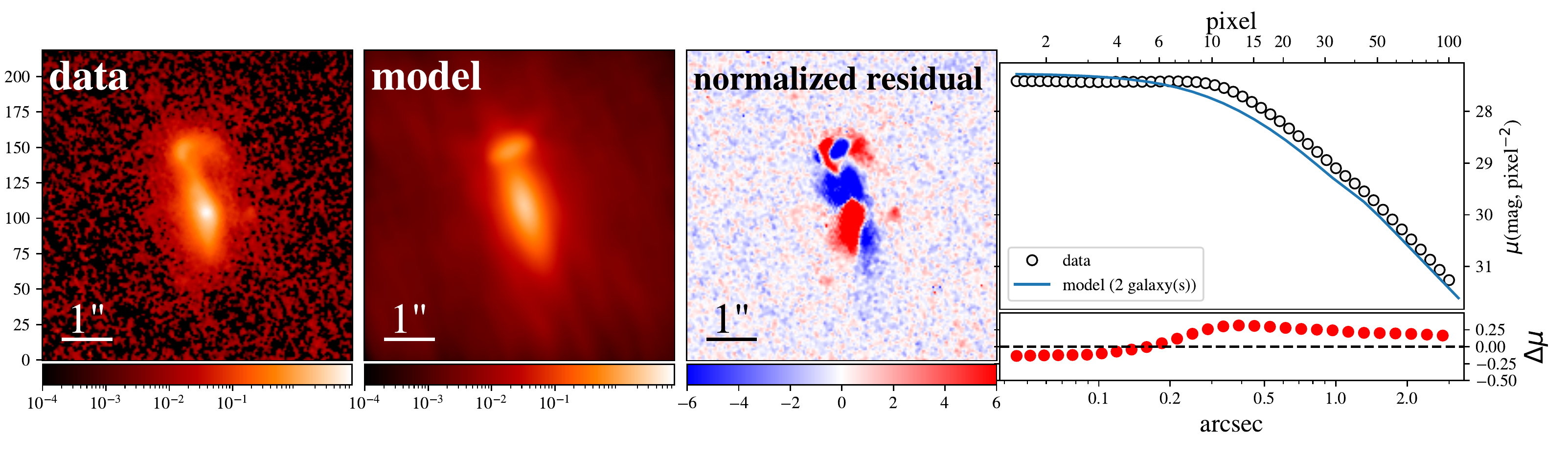}
\caption{2D decomposition of CID 473 using the F444W filter. The pixel scale is 0\farcs03\,pixel$^{-1}$. The panels are as follows: data, model, data minus point source (host galaxy only), normalized residual (data--model/$\sigma$), and surface brightness profile. The top panel includes an unresolved point-source component to the model while the bottom row does not. It is apparent that a central component improves the fit.}
\label{fig:example_decomp}
\end{figure*}

\subsection{X-ray AGN}

The $Chandra$ COSMOS Legacy survey \citep{Civano2016} provides an X-ray catalog and properties for 4016 point sources over 1.5\,deg$^2$ of the COSMOS field. The flux limits of the survey reach down to depths of $2.2~(8.9)\times10^{-16}$\,erg\,cm$^{-2}$\,s$^{-1}$ in the 0.5--2 (2--10)\,keV band, respectively. The X-ray obscured ($N_H>10^{21.5}$\,cm$^{-2}$) fraction is $\sim50\%$ of the AGN population over a wide range of X-ray luminosity \citep[e.g.,][]{Merloni2014}. We use the optical/infrared counterparts to X-ray sources as reported in \cite{Marchesi2016a} to identify the AGN host galaxies for further study with JWST. The X-ray positions are accurate to $\pm1^{\prime\prime}$ thus the optical associations are secure.

There are 50 X-ray AGNs from the \citet{Civano2016} catalog falling within these early 6 COSMOS-Web NIRCam pointings. Forty-six have reliable spectroscopic redshifts. Of these, there are 22 in the \citet{Marchesi2016a} catalog with $log~N_H>21.5$, our threshold for being labeled as obscured. In this work we focus on three AGN, namely CIDs 473, 1210, and 1245.

This catalog also provides X-ray column densities. These are computed from the X-ray hardness ratio assuming a photon index $\Gamma$=1.8. The hardness ratio is defined as $\mathrm{HR}=(H-S)/(H+S)$, where H and S are the net counts measured in the 2--7\,keV and 0.5--2\,keV bands, respectively. The hardness ratio is
an effective proxy of the line-of-sight column density, since with increasing $N_H$ values the 2--7\,keV observed emission becomes progressively more dominant with respect to the 0.5--2\,keV emission. As an example, in 0.5--7\,keV emission CID\,473 has more than 30 net counts thus a full spectral fit was performed in \citet{Marchesi2016b}. The measurement ($\log N_H=23.01_{-0.63}^{+0.31}$ cm$^{-2}$) is fully consistent with the one obtained using the HR, a result that supports the validity of the HR-derived measurements for the other two targets.

\begin{deluxetable*}{lllllllllll}
\tabletypesize{\scriptsize}
\tablecaption{AGNs in COSMOS-Web with galactic-scale obscuration\label{tab:sample}}
\tablehead{
\colhead{X-ray ID}&
\colhead{RA}&
\colhead{Dec}&
\colhead{Redshift\tablenotemark{a}}&
\colhead{$N_X$\tablenotemark{a}}&
\colhead{log\,$L_{X}$\tablenotemark{a}}&
\colhead{$N_{H}$\tablenotemark{a}}&
\colhead{$E(B-V)_{\rm host}$}&
\colhead{$E(B-V)_{\rm host}$\tablenotemark{b}}&
\colhead{$E(B-V)_{\rm host}$\tablenotemark{c}}&
\colhead{$E(B-V)_{\rm total}$\tablenotemark{b}}\\
&&&&&&($\times10^{22}$\,cm$^{-2}$)&(Global)}
\startdata
CID\_1245&149.997330&2.449278&0.759&15.6&42.96&10.8$^{+9.1}_{-3.5}$&0.54&0.98$\pm$0.17&0.88$\pm$0.08&$>1.27$\\
CID\_473&149.979492&2.309112&1.456&63.4&43.9&11.2$^{+5.1}_{-4.2}$&0.43&0.94$\pm$0.13&0.90$\pm$0.06&1.20$\pm$0.39\\
CID\_1210&149.953552&2.383046&1.924&22.2&43.7&$>35$&0.29&0.64$\pm$0.18&0.57$\pm0.08$&1.16$\pm$0.44\\
\enddata
\tablenotetext{a}{Redshift, X-ray counts (2--10\,keV), rest-frame luminosity (2--10\,keV), and absorbing column density as reported in \citet{Marchesi2016a}.}
\tablenotetext{b}{SED Fitting code: MICHI2}
\tablenotetext{c}{SED Fitting code: CIGALE}
\end{deluxetable*}

\section{Methods}

\subsection{2D AGN-host galaxy image decomposition}
\label{text:2d_decomp}
As part of a larger effort to study the host galaxies of known AGNs within the COSMOS-Web survey area, we decompose the JWST images of the 50 $Chandra$ X-ray sources having early NIRCam data into AGN and host galaxy components. This allows us to determine whether an extended host galaxy and/or an unresolved component -- i.e., an AGN -- are detected in each band and measure their properties (e.g., magnitude, size, profile shape) without a contribution from the other. The analysis is performed based on our self-developed software \galight\ \citep{Ding2021}, which is a \texttt{python}-based open-source package that provides various astronomical data processing tools and performs 2D profile fitting. It utilizes the image modeling capabilities of \lenstronomy\ \citep{Birrer2018} while redesigning the user interface to allow for an automated fitting ability. The AGN is modeled using an empirical point-spread function based on stars within the COSMOS-Web field-of-view while the host is modeled as a smooth \ss~profile. The FWHM of the PFS depends on the filter (F115W: $\sim0.053^{\prime\prime}$, F150W: $\sim0.062^{\prime\prime}$, F277W: $\sim0.125^{\prime\prime}$, F444W: $\sim0.163^{\prime\prime}$). We refer the reader to \citet{Ding2022,Ding2023} for further demonstrations of the 2D analysis of AGN and their host galaxies using JWST imaging.

In Figure~\ref{fig:example_decomp}, we show the 2D image decomposition of CID 473, one of the three X-ray AGN studied here. The pixel scale of the NIRCam image is 0\farcs03\,pixel$^{-1}$. Three components comprise the model to fit two galaxies and an AGN. The additional component accounts for the companion galaxy likely interacting with the AGN host. With the F444W image, an AGN component is detected with a magnitude of $21.47\pm0.08$. This value is determined using a PSF-model based on a single star which resulted in the lowest reduced $\chi^2$ for the fit. The uncertainty is then assessed by using five PSF models, each constructed from different stars. All further analysis is based on the decomposition using the best-performing PFS-model. While we are cautious in our claim of an unresolved point source or bulge component, it is evident that the fitting based on a model without an unresolved point source component performs far worse since the reduced $\chi^2$ increases to 10.3 from 4.4. In any case, consideration of the AGN magnitude as a lower limit in subsequent analysis does not detract from our main science as presented below. For the galaxies of interest here, the AGN component is primarily detected in the redder JWST bands and completely hidden in the bluer bands due to extinction. The 2D fits are shown for CID 1245 and CD 1210 in the appendix.

\begin{figure*}
\epsscale{0.95}
\plotone{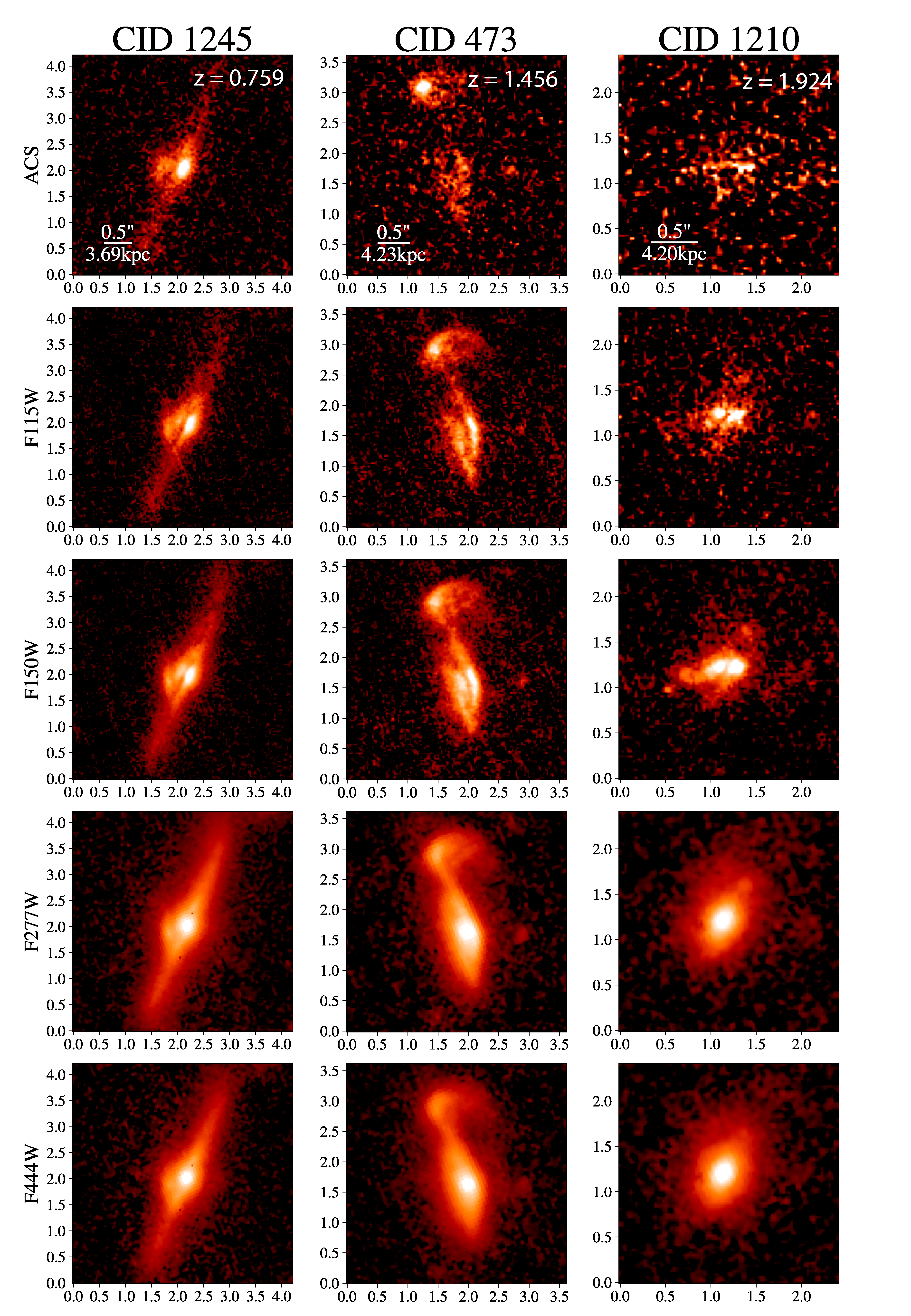}
\caption{HST/ACS F814W and JWST/NIRCam (F115W, F150W, F277W, F444W) images of three X-ray AGN in COSMOS-Web exhibiting galaxy-scale dust lanes. The axes are labeled in units of arcsecs while the physical scale is also shown in the top panels. The galaxies are ordered by increasing redshift (shown in the top panel) from left to right.}
\label{fig:cw_images}
\end{figure*}

\subsection{SED fitting}
\label{text:sed}
The above analysis produces an image of the host galaxy in each band free of any AGN emission to perform spatially-resolved SED fitting over the central region (i.e., the dust lane; see below). As a caveat, we note that any mismatch of the PSF in the modeling may affect the colors (i.e., $E(B-V)$) of the inner regions of the host galaxy. We use the codes \texttt{gsf} \citep{Morishita2019}, \texttt{MICHI2} \citep{Liu2021} and \texttt{CIGALE} \citep{Boquien2019} to perform SED fitting to the JWST photometry. The former provides an application for spatially-resolved images with ease while the latter two include the AGN templates of interest. For the host galaxy, stellar templates are from \citet{Bruzual2003} with solar metallicity, constant star formation history, and a Chabrier IMF \citep{Chabrier2003}. For CIGALE, the constant star formation history has a timescale of 200 Myr whereas MICHI2 is taken over cosmic time. The age is a free parameter which is poorly constrained to be less than a Gyr. For the AGN photometry, we use the \citet{Lyu2017} quasar template. For both sets of photometry, quasar and host galaxy, an attenuation law is applied assuming $R_V=3.1$. Given the limited photometric bands, the inclusion of additional SED components such as the thermal dust emission is not necessary. Our interest is to derive likelihood distributions for the reddening $E(B-V)$ which MICHI2 provides; these are used throughout the following quantitative analysis including the uncertainty on $E(B-V)$.

\section{Results}

We visually inspected JWST/NIRCam cutout images of 50 X-ray AGNs covered by the available COSMOS-Web imaging. Both the raw science frames and images with the best-fit model AGN removed (i.e., host galaxy only) were available. The vast majority of the AGN have detections of their host galaxies in exquisite detail not previously seen for deep X-ray selected samples. The full results from this analysis will be presented in a subsequent study.

Of these 50 X-ray AGN, three stood out for clearly exhibiting structures characteristic of dust lanes in the bluer bands (F115W and F150W) which have not been detected in HST imaging for AGNs at $z\gtrsim1$. Their identification did not require 2D decomposition since the AGN signal is very weak (see below). Due to the wealth of spectroscopy in the COSMOS field, these three AGNs have spectroscopic redshifts of $z=0.759$, 1.456 and 1.924 as reported in \citet{Marchesi2016a}. These cases represent 27\%\footnote{There is a fourth case, CID 530 ($z_{spec}=0.931$), which shows hints of the presence of a dust lane (not shown). If realized, then 36\% of the heavily obscured ($N_H\gtrsim10^{23}$\,cm$^{-2}$) population would have a significant contribution of the X-ray absorption attributed to the host galaxy} (3 out of 11 falling within the early COSMOS-Web NIRCam images) of the highly obscured ($N_H\gtrsim10^{23}$\,cm$^{-2}$) population.

It is now apparent that JWST imaging can elucidate the nature of the obscuring material. A galactic component to the extinction is likely contributing to the nuclear obscuration which effectively hides an embedded AGN in each case, especially for the bluer bands. The total photometry for these three sources is predominantly due to the host galaxies as evident by their IRAC colors; they do not satisfy the IR color selection for AGNs \citep{Donley2012}. In Table~\ref{tab:sample}, we list their properties.

In Figure~\ref{fig:cw_images}, we show image cutouts of the three galaxies in the four NIRCam bands plus the original COSMOS HST/ACS F814W band \citep{Koekemoer2007,Scoville2007}. The impact of dust is present in the F115W and F150W images while the redder bands are at wavelengths long enough for the emission to penetrate through the obscuration. For instance, both CID-473 ($z=1.456$) and CID-1245 ($z=0.789$) have a prominent dust lane likely the result of the galaxy being viewed nearly edge-on. In these two cases, the dust lanes extend along the plane of the galaxy and have a scale height of 0.9 and 0.5\,kpc, respectively. In the top panels of Fig.~\ref{fig:cw_images}, the spatial scale for each case is given. CID 473 appears to be in a merger with a nearby galaxy to the north, likely contributing to its more complex morphology. The most distant, CID-1210 ($z=1.924$), has two separate sources in the bluer bands which is a single source in the redder bands. We note that the position angle of the elongation in the F$444$W band may signify the direction of a dust lane which is not at the same position angle as the two separate sources seen in F115W. Considering the COSMOS HST/ACS F814W images, we note that two of the sources are at sufficiently high redshift (CID-473 and CID-1210) such that these images sample the rest-frame SED at near-UV wavelengths, well below the 4000 $\AA$ break, thus lacking sufficient stellar continuum emission to clearly reveal the properties of the dust lanes.

\begin{figure*}
\epsscale{1.1}
\plotone{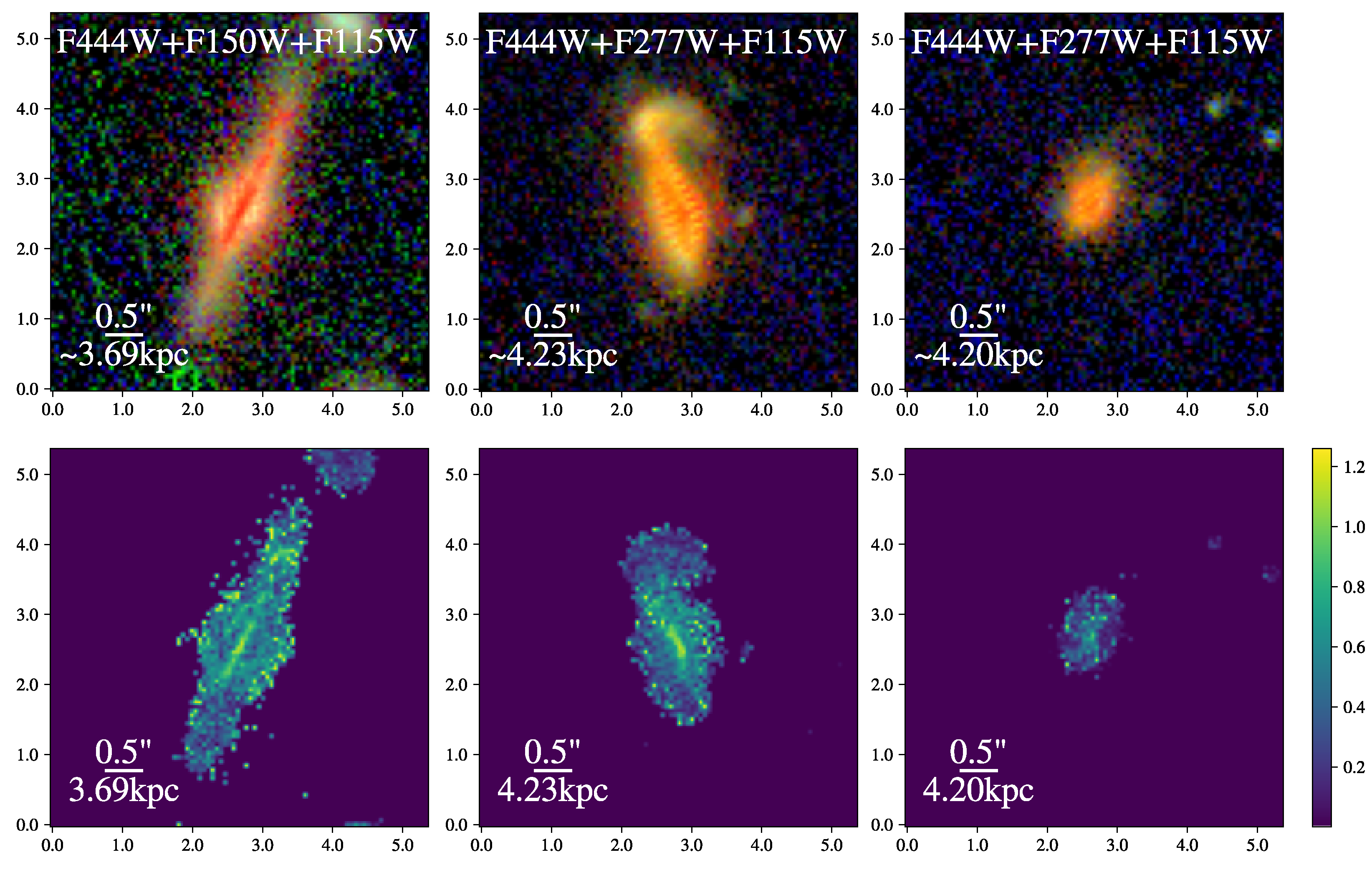}
\caption{Color images (RBG, top panels) generated by using the F444W, F277W and F115W filters. Reddening maps $E(B-V)$ are shown in the bottom panels which are binned by a factor of two thus have a pixel scale of 0\farcs06. The axes of all images are given in arcsecs.}
\label{fig:ebv}
\end{figure*}

\begin{deluxetable}{llll}
\tabletypesize{\scriptsize}
\tablecaption{2D image decomposition\label{tab:decomp}}
\tablehead{
\colhead{X-ray ID}&
\colhead{Filter}&
\colhead{Host mag}&
\colhead{AGN mag}}
\startdata
CID\_473&F115W&22.73$\pm$0.07&$>24.29$\\
&F150W&21.95$\pm$0.02&$>24.81$\\
&F277W&20.98$\pm$0.03&22.62$\pm$0.09\\
&F444W&20.45$\pm$0.04&21.47$\pm$0.37\\
CID\_1210&F115W&23.69$\pm$0.01&$>$27.30\\
&F150W&22.53$\pm$0.08&$>26.26$\\
&F277W&21.99$\pm$0.01&25.28$\pm$0.12\\
&F444W&21.59$\pm$0.05&24.13$\pm$0.37\\
CID\_1245&F115W&21.53$\pm$0.01&$>$23.21\\
&F150W&20.79$\pm$0.02&$>$23.78\\
&F277W&19.74$\pm$0.01&$>$22.35$\pm$16\\
&F444W&20.12$\pm$0.02&22.56$\pm$0.23\\
\enddata
\end{deluxetable}

\begin{figure}
\epsscale{1.0}
\plotone{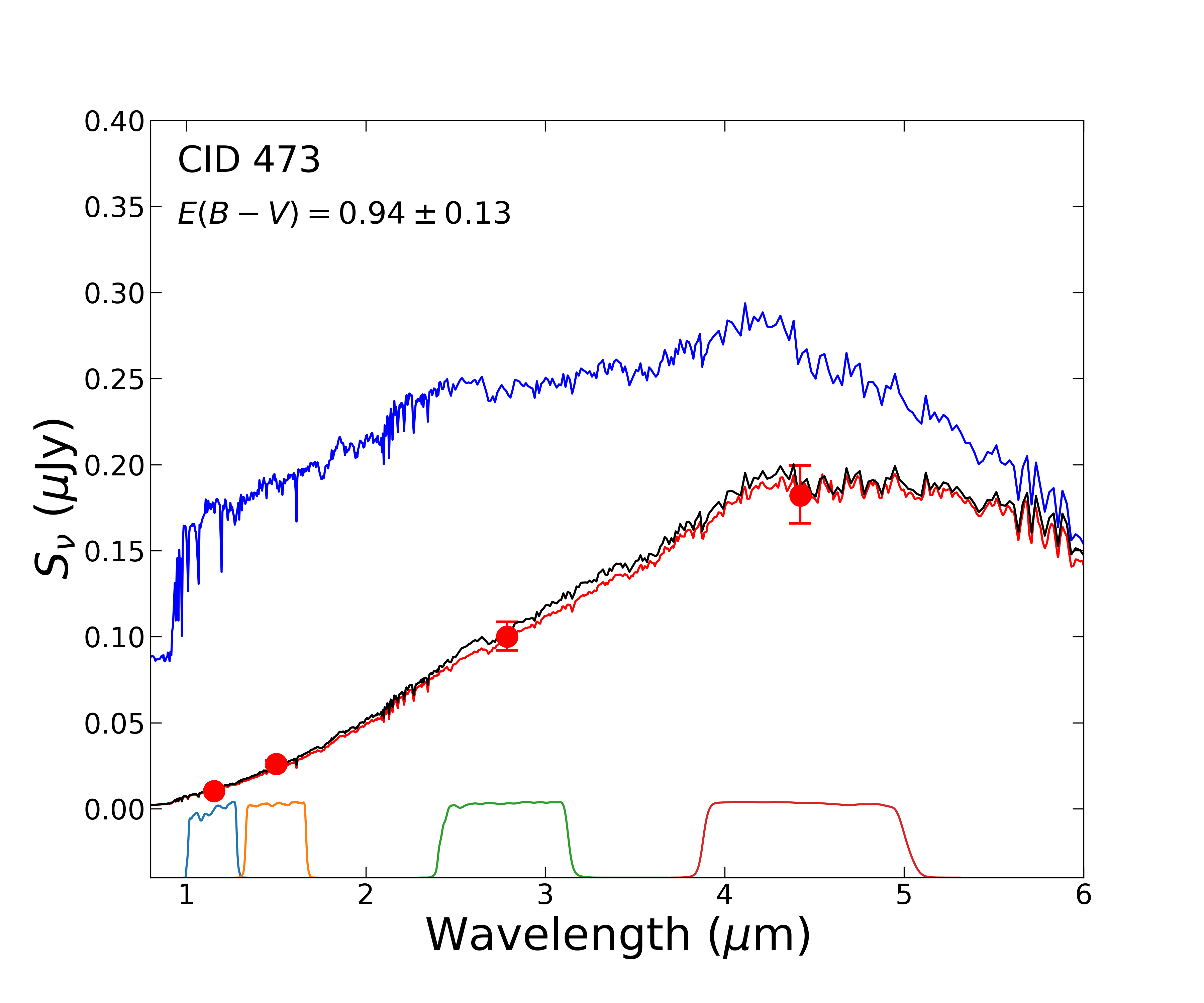}
\plotone{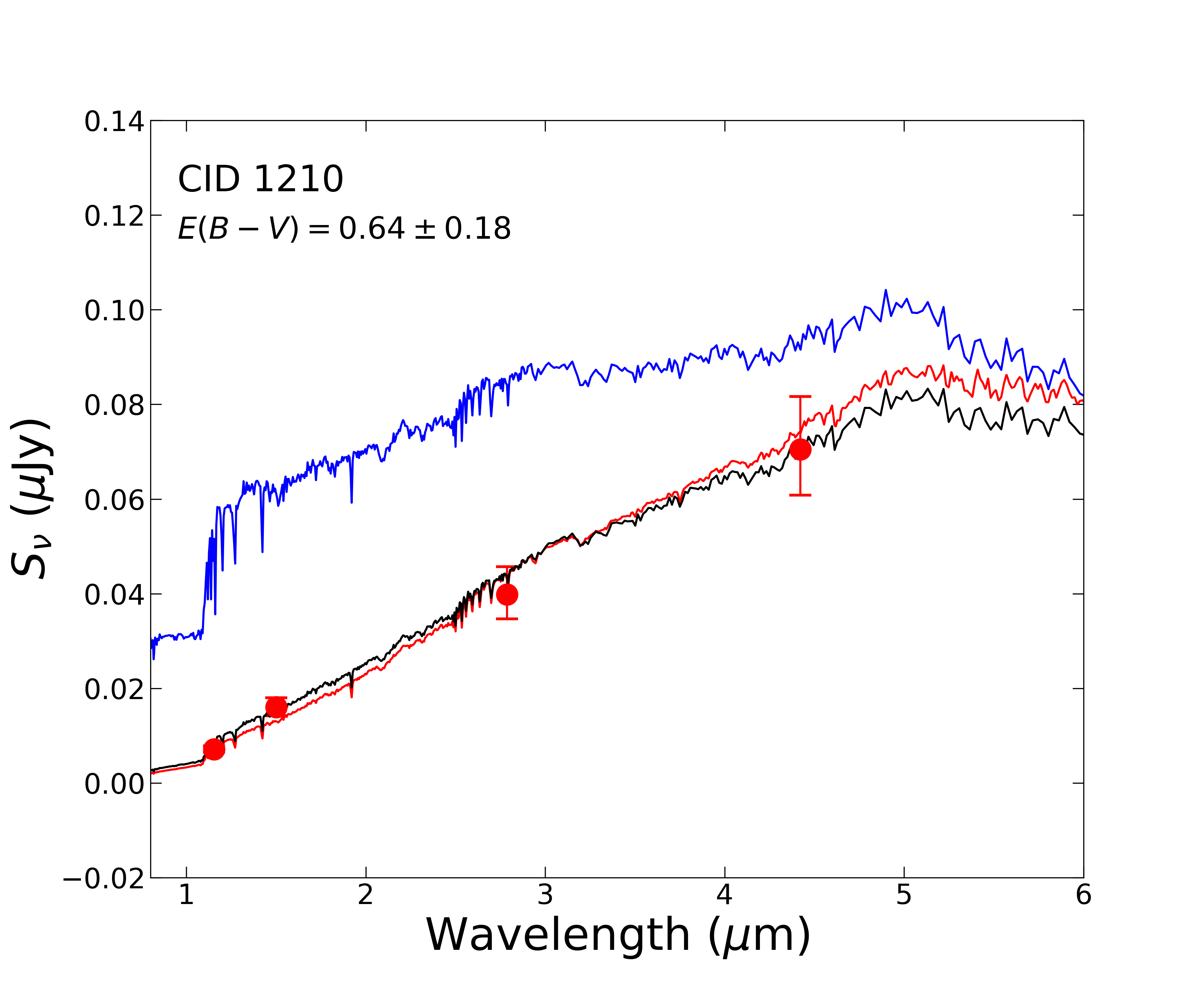}
\plotone{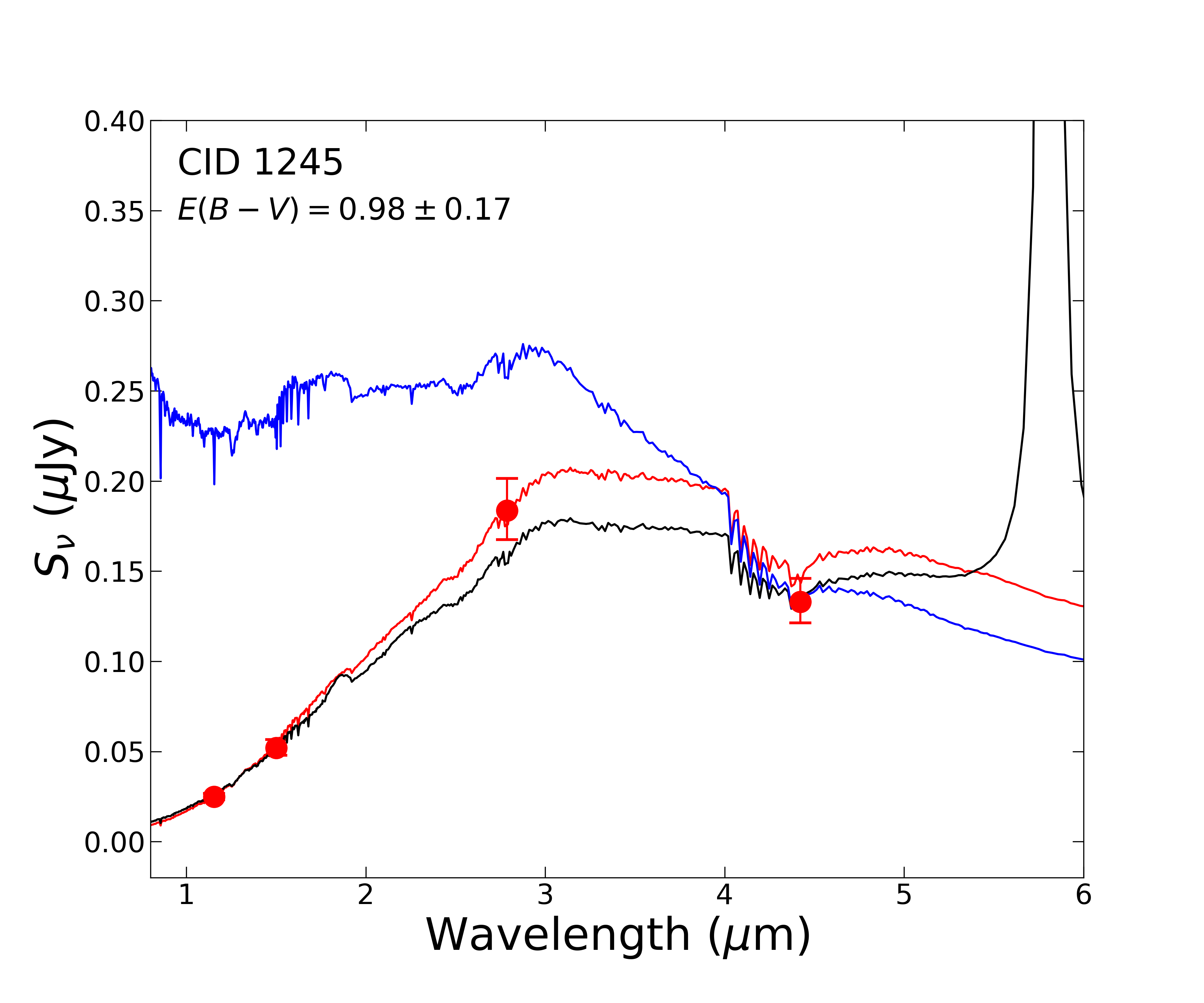}
\caption{Host galaxy fluxes and best-fit SEDs using MICHI2 (red) and CIGALE (black). The observed fluxes are given in red with 1$\sigma$ uncertainties. The unattenuated model SED from CIGALE is shown in blue. The JWST/NIRCam filters are shown in the top panel.}
\label{fig:galaxy_sed-fits}
\end{figure}

\subsection{Spatially-resolved extinction maps of the host galaxies}

 In Figure~\ref{fig:ebv}, the results are shown for a pixel-by-pixel modeling of the spectral energy distribution of the JWST images to estimate the color excess $E(B-V)$ attributed to the host galaxy. To improve the significance, we bin the images by a factor of two. The $E(B-V)$ estimates are shown only for spatial elements with signal-to-noise greater than 5. For comparison, we show the color composite images for each. 
 
 While our 2D decomposition finds marginal evidence for an AGN in the F277W (F444W) bands for 2 (3) cases (Table~\ref{tab:decomp}), the AGN contributes insignificantly to the total near-infrared emission. Hence, the light is primarily due to the stellar component; thus $E(B-V)$, measured with the original science frames, is attributed to the host galaxy, assuming an adequately-matched intrinsic galaxy SED (see Sec.~\ref{text:sed}), with no contribution from nuclear dust (i.e., a torus). Therefore, we do not need to remove the AGN for this exercise. 
 
 As expected, we find that $E(B-V)$ is highest along the dust lanes with values of 0.6--0.9 (Cols 9 and 10 in Table~\ref{tab:sample}) at the location of the AGN where the IR emission peaks. In Figure~\ref{fig:galaxy_sed-fits}, we show the SED fits to the emission within a rectangular region with the center and dimension set to cover the extent of the dust lanes. Based on these, we report the mean value of $E(B-V)$ in each case (Table~\ref{tab:sample}). With global estimates of $E(B-V)$ being lower ($0.3 \lesssim E(B-V)\lesssim0.5$; Table~\ref{tab:sample}), the high spatial resolution observations with JWST has enabled a more accurate assessment of the obscuration along the line-of-sight to the embedded AGN. However, it is possible that the actual value of the obscuration to the AGN, on scales smaller than resolved by JWST, may be different than the average value. Even so, we use these average estimates of the reddening to assess their likely contribution to the total X-ray absorbing column density toward the AGN.

\begin{figure}
\epsscale{1.0}
\plotone{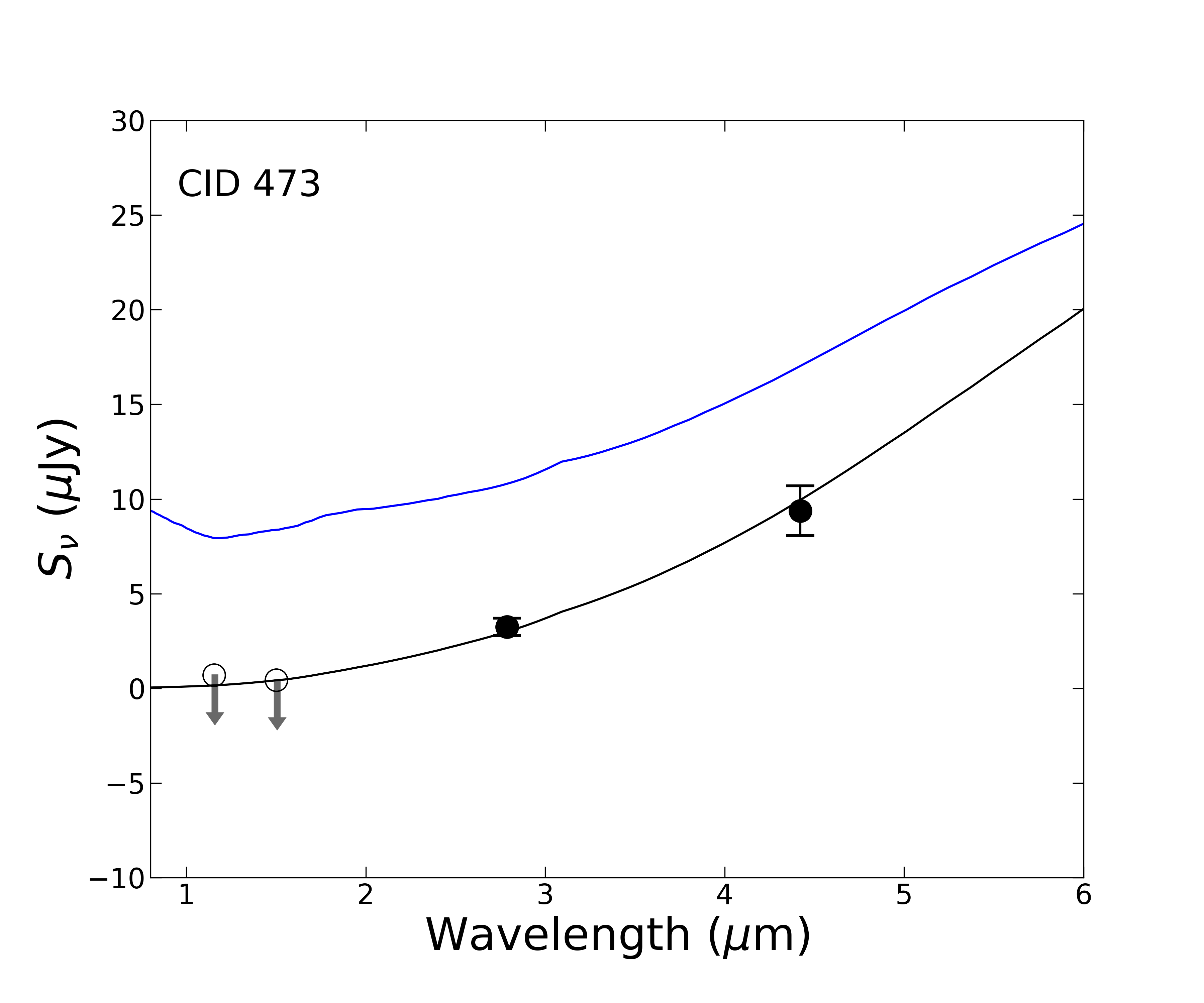}
\plotone{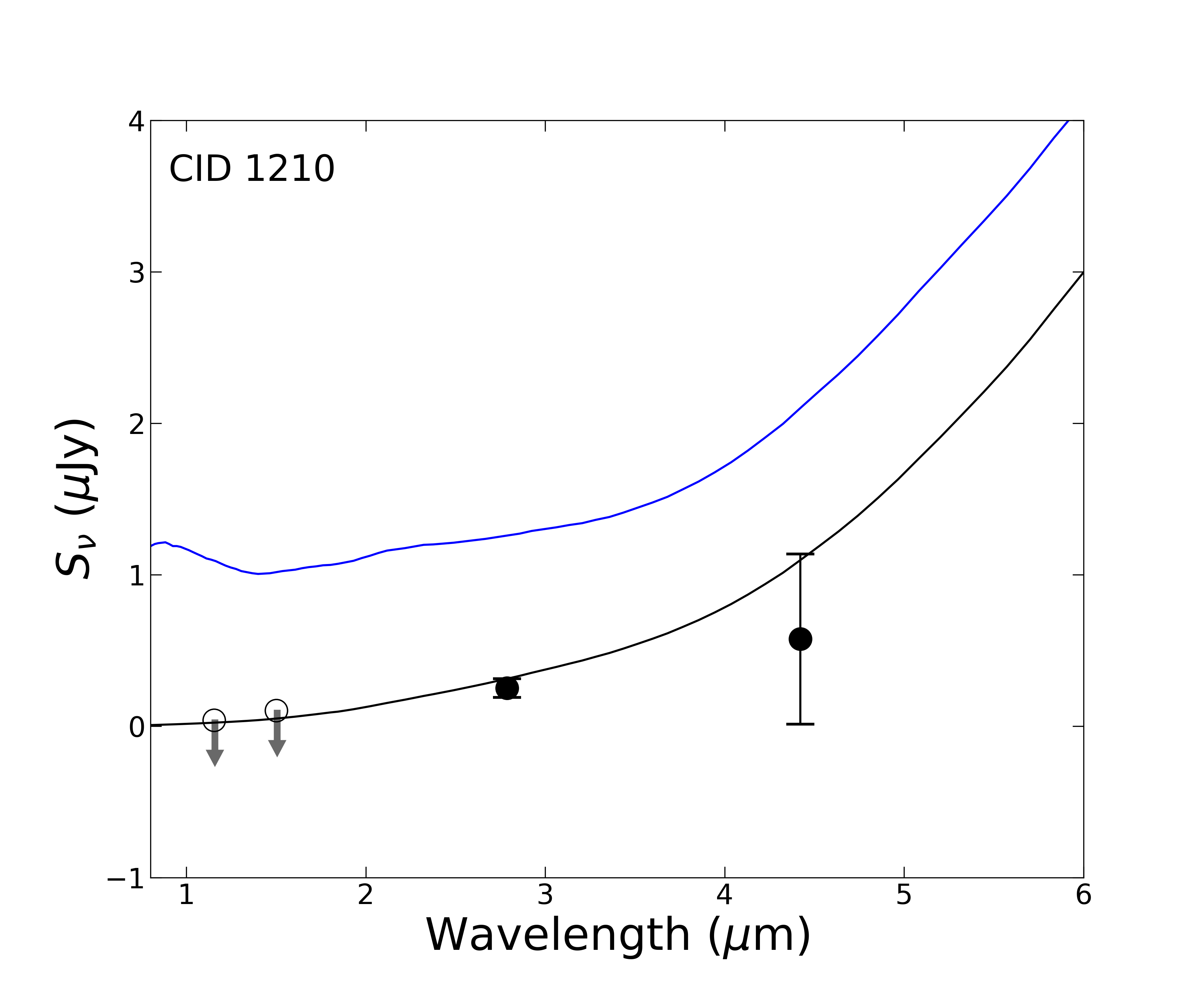}
\plotone{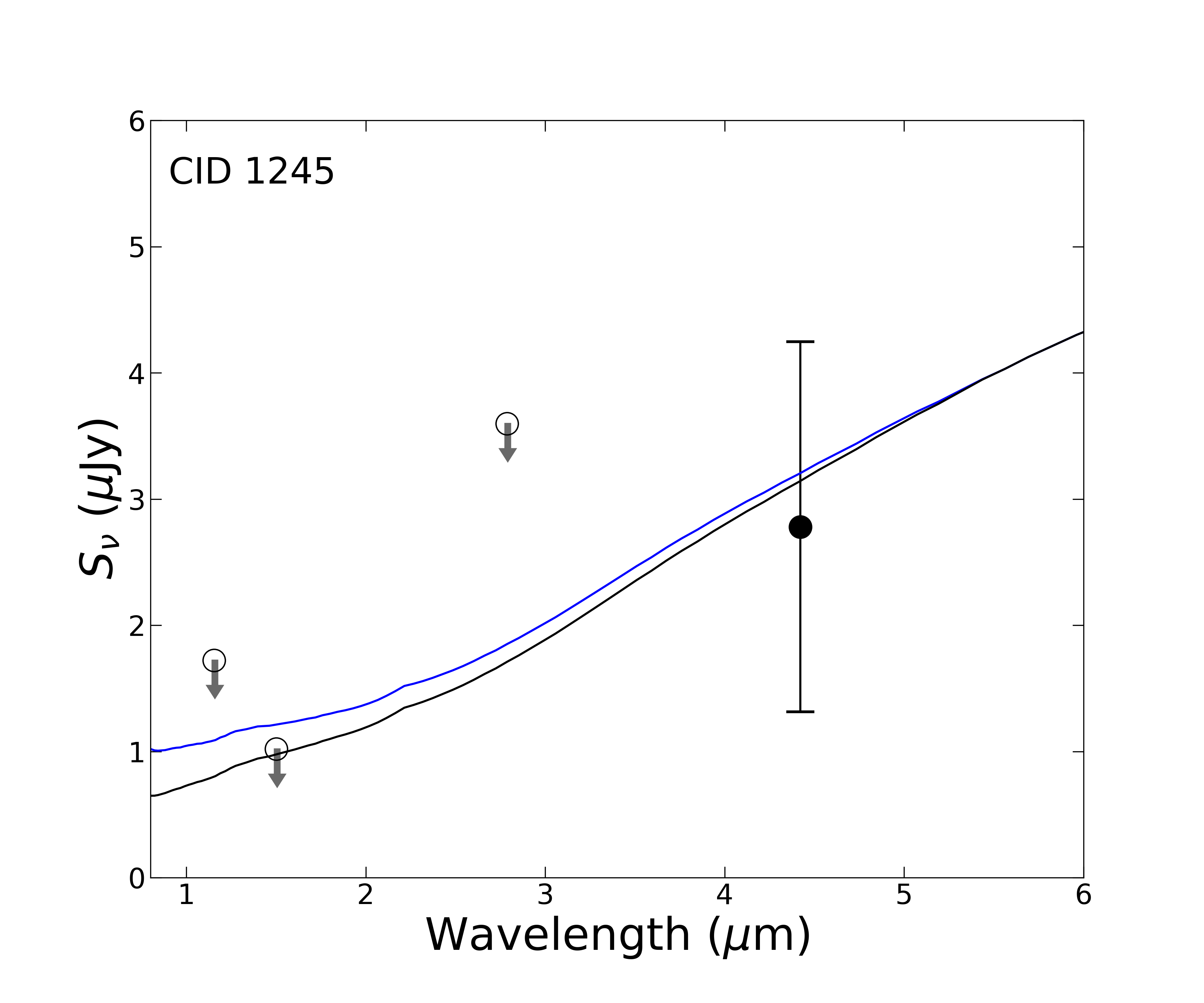}
\caption{Decomposed AGN JWST fluxes (circles) and best-fit SED (blue=unattenuated; black=attenuated) from \citet{Lyu2017}. Open symbols indicated upper limits on AGN emission.}
\label{fig:agn_sed-fits}
\end{figure}

\subsection{AGN optical obscuration}

Based on the AGN photometry from our 2D decomposition (Table~\ref{tab:decomp}), we assess the level of extinction possibly obscuring an embedded AGN in these three cases. In Table~\ref{tab:decomp}, we provide the photometry of the AGN component in all four NIRCam bands. The AGN magnitude is determined by summing all pixel values in the 2D-model AGN component within the image cutout. An AGN component is detected in the reddest band (F444W) for each case. For two, an AGN component is also detected with F277W. The other bands provide upper limits which constrain the level of extinction.

As a simple exercise, we fit the photometry using the \citet{Lyu2017} intrinsic quasar template in each case while varying $E(B-V)$ to match the observed data (Figure~\ref{fig:agn_sed-fits}). In the two cases with detections in both the F277W and F444W bands, we are able to measure considerable levels of extinction (CID-473: $E(B-V)=1.20\pm0.39$; CID-1210: $E(B-V)=1.16\pm0.44$). For CID-1245, the constraint is weaker ($E(B-V)>1.27$) given the single band detection. We note that the values here represent the total level of extinction along the line-of-sight to the AGN which includes the host galaxy contribution. We recognize that CID 1245 has very poor constraints given the photometric uncertainties and upper limits thus we may no claim to detecting the rest-frame optical emission from the AGN. Survey fields with deeper NIRCam imaging may be more effective at detecting the AGN in similar systems across multiple bands (see Sec.~\ref{text_ceers}).

\begin{figure}
\epsscale{1.3}
\plotone{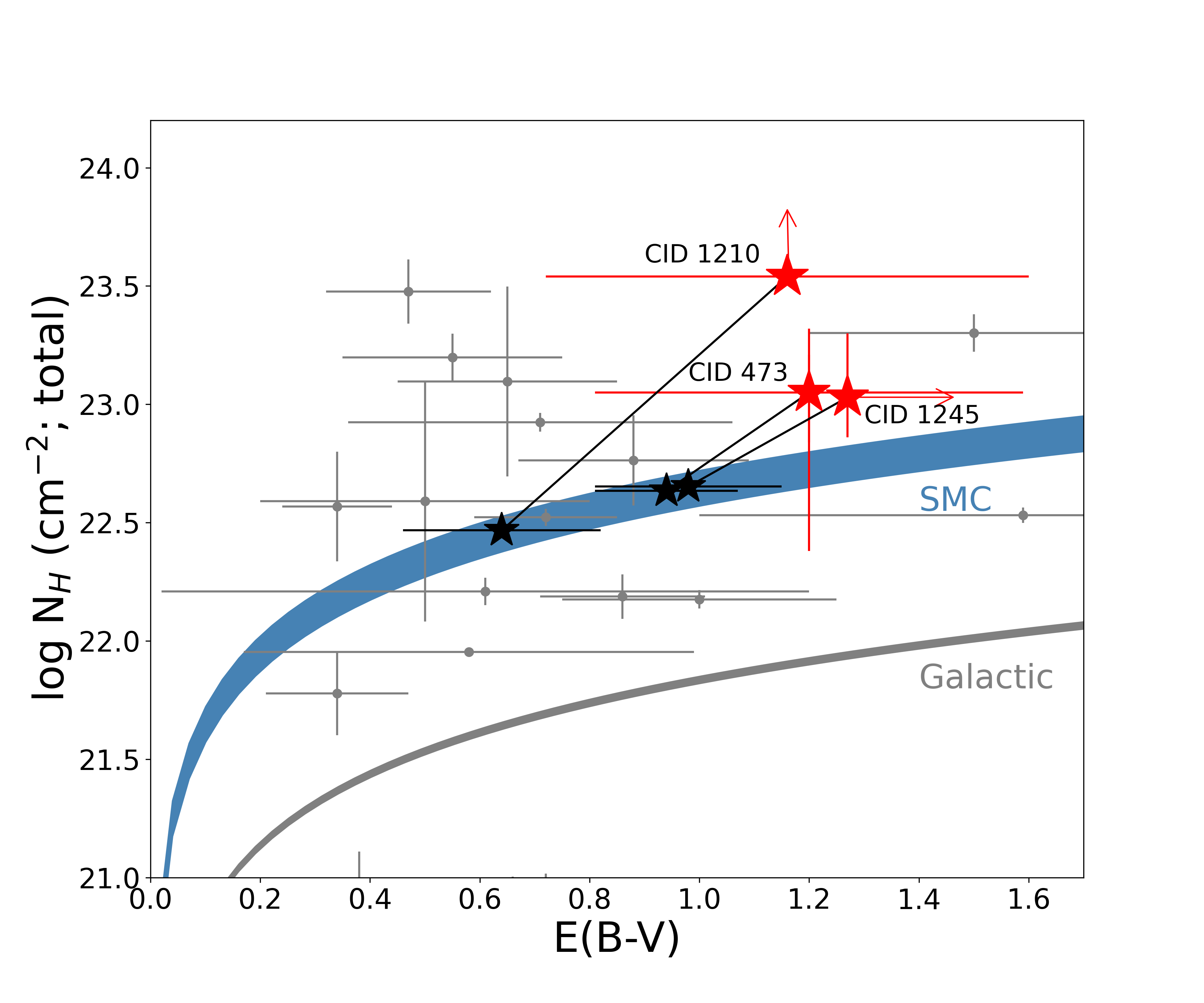}
\caption{Total X-ray absorbing gas column ($N_H$) versus dust reddening $E(B-V)$ for the host galaxy and AGN. The contribution to the obscuration from the host galaxy is shown by the black stars; their $N_H$ values are those expected based on the average central values of $E(B-V)$ and the SMC gas-to-dust ratio as indicated by the shaded blue region. The $N_H$ contribution from the host could be lower if the Galactic gas-to-dust ratio (in grey) is assumed. For CID 473, CID 1210 and CID 1245, their total obscuration is indicated by the red stars. These include the contribution of the host galaxy and inner obscuring structure (i.e., torus). For comparison, we plot a local AGN sample \citet{Maiolino2001} in grey.}
\label{fig:ebv_nh}
\end{figure}

\subsection{Optical vs. X-ray obscuration}

We use our extinction measurements to assess the contribution of the host galaxy to the X-ray obscuring column density of neutral gas. In Figure~\ref{fig:ebv_nh}, we plot the X-ray absorption $N_H$ as a function of $E(B-V)$. As previously mentioned, our three X-ray AGN are highly absorbed with column densities $N_H\gtrsim10^{23}$\,cm$^{-2}$. For CID 473, we plot the column density based on the direct spectral fit while the other two are based on the hardness ratios thus their uncertainties are likely to be larger than shown for CID 1210 and CID1245. We further note that additional X-ray spectral components \citep{Fabbiano2018} not accounted for may lead to additional uncertainties on these column densities. For the three AGN components having $E(B-V)$ measurements including the one with a lower limit, we mark their location with a red star symbol. For reference, we display the typical gas-to-dust relation ($N_H/E(B-V)$ between 3.7--5.2$\times10^{22}$\,atoms\,cm$^{-2}$\,mag$^{-1}$) based on the SMC \citep{Bouchet1985} which is applicable for X-ray obscured AGN \citep{Maiolino2001, Willott2004}. In addition, we indicate the Galactic gas-to-dust relation \citep{Guver2009} which would result in lower column density contributions of the host. We include the local sample of Seyfert galaxies from \citet{Maiolino2001} to demonstrate that our three cases are elevated in their X-ray column densities relative to their optical extinction, similar to well-known low redshift AGNs.

For all three, we show their expected value of $N_H$ given their measured $E(B-V)$ of the host (black symbols). It is clear that the extinction due to the host galaxy can account for $N_H\sim10^{22.5}$\,cm$^{-2}$ assuming that the measured gas-to-dust ratio of the SMC is applicable. This amounts to a host galaxy contribution of $\lesssim$\,30\% of the total X-ray column density. These values are in very good agreement with the median ISM column densities, $<N_H>\sim10^{22}-10^{22.5}$\,cm$^{-2}$  , reported between z$\sim1-2$ in \citet{Gilli2022}. The remaining obscuring component is then attributed to a central component likely an obscuring torus. As one consequence, these estimates of the X-ray absorbing column density due to the host galaxy may further limit the size of the forbidden zone for AGN in the plane of $N_H - \lambda_{Edd}$ \citep{Fabian2009} thus making it more challenging to link radiation pressure to the lack of obscuring clouds in galactic nuclei.

\begin{figure*}
\epsscale{1.0}
\plotone{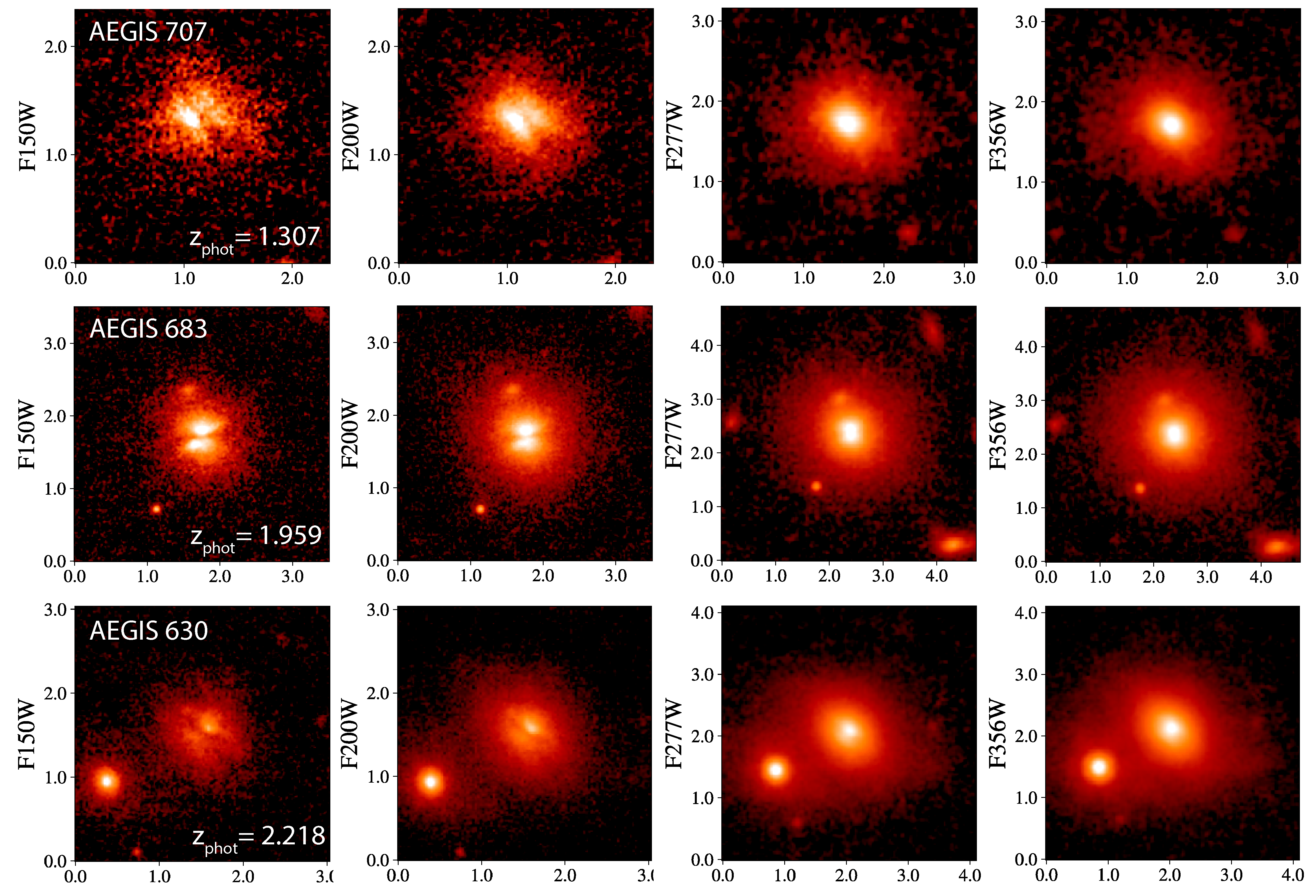}
\caption{X-ray AGNs in CEERS with prominent galaxy-scale dust lanes in the bluer filters (F150W and F200W). Each row displays the individual JWST images for each filter for a single source. Photometric redshifts are as labeled. The axes are given in arcsecs.}
\label{fig:ceers}
\end{figure*}

\subsection{Additonal cases in CEERS}
\label{text_ceers}
As a demonstration of their proliferation, similar cases are also present in the JWST ERS program CEERS \citep{Finkelstein2022}. We searched the catalog of X-ray AGN from the \textit{Chandra} imaging \citep{Laird2009,Nandra2015} of the AEGIS survey, i.e., the central region of the Extended Groth Strip. The X-ray data reaches down to limiting fluxes of $1.5~(2.5)\times10^{-16}$\,erg\,cm$^{-2}$\,s$^{-1}$ in the 0.5--2 (2--10)\,keV band. Out of 59 X-ray sources with NIRCam imaging, we find three additional cases (AEGIS 707, 683 and 630) with prominent dust lanes. These have photometric redshifts of 1.307, 1.959 and 2.218 respectively. Their JWST images in four filters are displayed in Figure~\ref{fig:ceers}.

Here, we highlight an exquisite case, AEGIS 683, with a narrow dust lane slicing across the central emission in the East-West direction in the bluer bands (e.g., F150W and F200W). In the redder bands (e.g., F356W), the dust lane disappears. The central emission in the long wavelengths bands is dominated by the galaxy with no signs of an optical AGN. Furthermore, the overall morphology of AEGIS 683 is symmetric and not elongated as would be expected for an inclined galaxy with such a prominent dust lane, as seen in AEGIS 707  and 630. AEGIS 683 appears to be a high-redshift analog to the closest radio galaxy Centaurus A which is thought to be the result of a merger of a gas-rich spiral galaxy and an elliptical galaxy. This source and the others presented herein highlight the remarkable galaxies to come from JWST in the study of the content and spatial distribution of obscuration in AGN and the non-active populations up to $z\sim2$.

\section{Conclusions}

We demonstrate the capability of JWST, using early NIRCam observations from the COSMOS-Web program, to resolve dust features in three galaxies up to $z\sim2$ which harbor obscured AGN. Prior to JWST, such structures were only able to be resolved for galaxies in the low redshift Universe \citep[e.g.,][]{Malkan1998,Bianchi2007,Juneau2022,Wu2022}. These structures are in the form of dust lanes evident from the edge-on nature of these selected X-ray AGNs. This population is not necessarily in the minority; these three cases represent 27\% of the 11 highly obscured ($N_H\gtrsim10^{23}$\,cm$^{-2}$) AGN observed within the first six NIRCam pointings. This supported by similar cases found in CEERS.

For the COSMOS-Web AGN, we generated spatially-resolved reddening maps to assess the level of obscuration to the AGN and that attributed to the host galaxy separately. Using a 2D image decomposition method, the extinction due to the host and that along the line-of-sight to the central AGN can be assessed. We find that the host galaxy has reddening $E(B-V)\sim0.6$--0.9 which contributes significantly to the total X-ray absorption, a few times $10^{22}$\,cm$^{-2}$, assuming the measured gas-to-dust ratio of the SMC. These levels of absorption are in agreement with that expected based on the demographics of AGN population \citep{Buchner2017}. In each X-ray AGN presented here, additional nuclear obscuration is required to account for the X-ray column densities of $\gtrsim10^{23}$\,cm$^{-2}$. These initial results are just a start to the progress to be achieved with furthering our understanding of the demographics of the AGN population and their host galaxies. By simply scaling these three cases to the final area to be observed with COSMOS-Web, we expect to have a sample of $\sim$75 similar cases with NIRCam imaging and $\sim$25 of these with the additional MIRI images when the full COSMOS-Web data set is available.

\section*{Acknowledgements}

We thank the anonymous referee for their comments which improved the paper. We also are appreciative of useful discussions with Andy Goulding and Rosie Wise. J.S. is supported by JSPS KAKENHI (JP22H01262) and the World Premier International Research Center Initiative (WPI), MEXT, Japan. M.O.\ is supported by the National Natural Science Foundation of China (12150410307). X.D.\ is supported by JSPS KAKENHI Grant Number JP22K14071. The Cosmic Dawn Center (DAWN) is funded by the Danish National Re- search Foundation (DNRF) under grant No. 140. B.T. acknowledges support from the European Research Council (ERC) under the European Union's Horizon 2020 research and innovation program (grant agreement number 950533) and from the Israel Science Foundation (grant number 1849/19). MF acknowledges support from NSF grant AST-2009577 and NASA JWST GO Program 1727. EV acknowledges support from Carl Zeiss Stiftung with the project code KODAR.






\appendix

\begin{figure*}
\epsscale{1.1}
\plotone{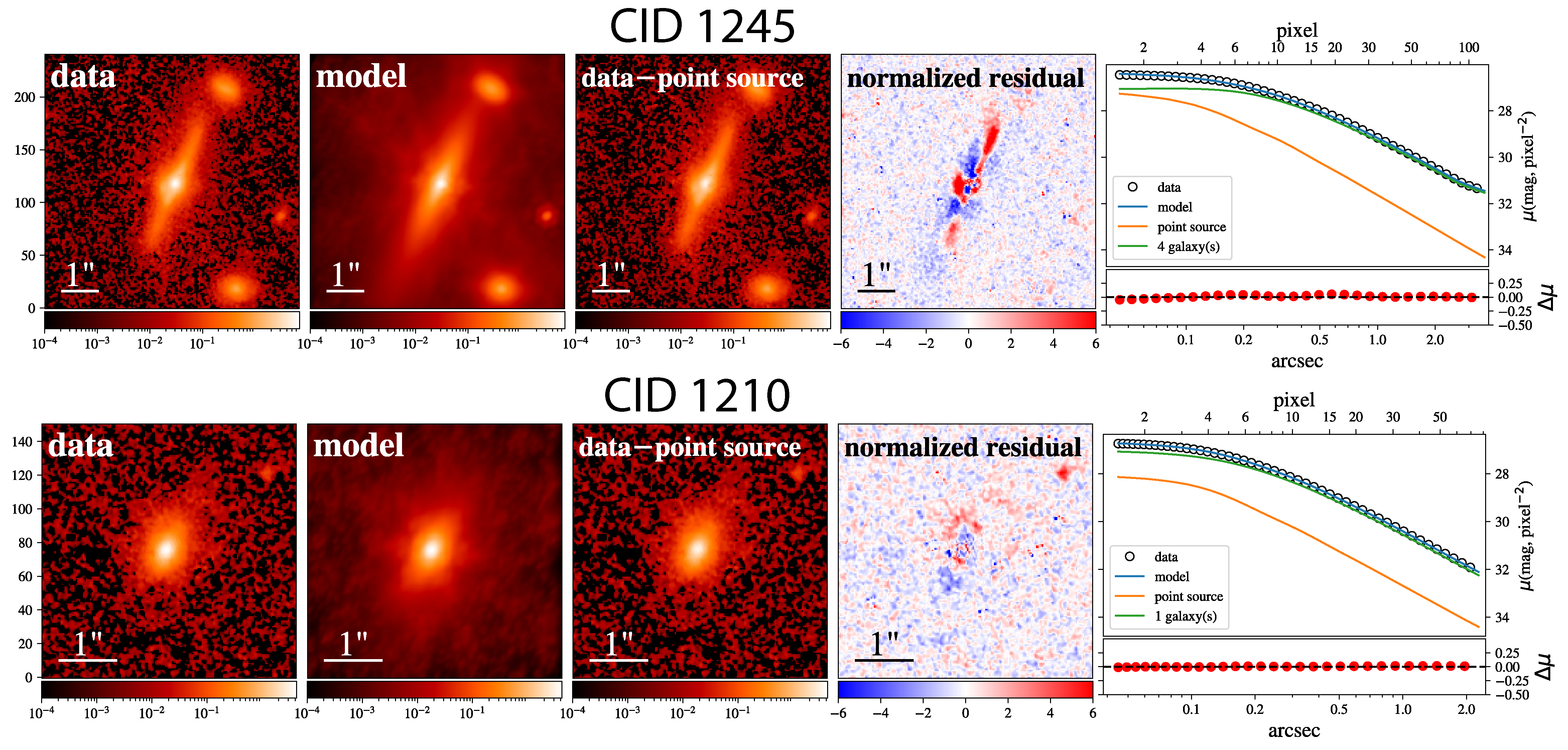}
\caption{2D decomposition of CID 1245 and CID 1210  using the F444W filter. The pixel scale is 0\farcs03\,pixel$^{-1}$. The panels are as follows: data, model, data minus point source (host galaxy only), normalized residual (data--model/$\sigma$), and surface brightness profile. All cases include an unresolved point-source component to the model.}
\label{fig_app:example_decomp}
\end{figure*}

As described in Section~\ref{text:2d_decomp}, the 2D model fitting of the JWST images is performed for CID 1245 and CID 1210. The best-fit model for each, including and AGN component, is shown in Figure~\ref{fig_app:example_decomp}.

\bibliography{jdsrefs}{}
\bibliographystyle{aasjournal}




\end{document}